\documentclass{article}
\pdfoutput=1

\AtBeginDocument{%
  \providecommand\BibTeX{{%
    \normalfont B\kern-0.5em{\scshape i\kern-0.25em b}\kern-0.8em\TeX}}}

\usepackage{graphicx}
\begin{document}

\title{X-Ray bone abnormalities detection using MURA dataset }

\author{Anna Solovoya, Igor Solovyov, 
TraumAI}

\begin{abstract}
  
 We introduce the deep network trained on the MURA dataset from the Stanford University released in 2017. Our system is able to detect bone abnormalities on the radiographs and visualise such zones. We found that our solution has the accuracy comperable to the best results that have been achieved by other development teams that used MURA dataset, in particular the overall Kappa score that was achieved by our team is about 0.942 on the wrist, 0.862 on the hand and o.735 on the shoulder (compared to the best available results to this moment on the official web-site 0.931, 0.851 and 0.729 accordingly). However, despite the good results there are a lot of directions for the future enhancement of the proposed technology. We see a big potential in the further development computer aided systems (CAD) for the radiographs as the one that will help practical specialists diagnose bone fractures as well as bone oncology cases faster and with the higher accuracy. 
\end{abstract}

\providecommand{\keywords}{image analysis, image interpretation and understanding, x-ray fracture detection, abnormality}

\maketitle

\section{Introduction and Motivation}

Muskuloskeletal traumas affect more than 1.7 billion people a year. Missing or misdiagnosing pathology after X-Ray or other radiology examination is the widespread problem that directly connected with human-factor as a result of huge workload due to increased number of patients and shorten the time that practical specialist has to evaluate the radiographs. These mistakes have serious consequences and often result in delayed treatment, worse long-term prognosis, increase of treatment costs and time, and can lead to disability of patients.  
The success of detection of the bone abnormalities combines both specialists experience and quality of the image. And while the quality of the equipment that are used for the X-ray imagine steadily improves, human factor in the bone fracture detection remains today an unsolved problem. There are a lot of factors that lead to the incorrect diagnosis, such as timing, level of experience of the radiologist and personal issues that might affect the quality of the work. For example, according to the data from the emergency departments there is a common peak in the biggest level of the radiographs interpretation error between 8 pm and 2 am (47 percent error in compare to 20 percent in the control group) \cite{Hallas}. Another problem that lead to the misdiagnosis is a big proportion of the young specialists, who lack the experience to detect the complicated injuries and usually need assistance from the more senior specialists to make the final diagnosis. Moreover, there are some areas of the skeleton, where detection of the abnormalities is very hard due to the limitation of the X-ray equipment and the lack of time to detect the right bone position of the image classification. The opposite situation when every x-ray image or other imaging examinations are assessed by 2-3 and more specialists has a negative effect on the speed and quality of the diagnostics. All experienced medical doctors in all specialities rely on radiologists and described findings, but always need to analyse all images themselves for the purpose of better understanding and better planning of treatment. Without systems that help and, for instance, highlight pathologie, each specialist does the same work and  image-assessment from the beginning .

Understanding all the challenges that face the clinical radiology branch today, our team decided to build the simple computer-aided diagnose system (CAD) to detect bone fractures on different parts of the human body and, thus, reduce the workload for the practical specialists and reduce the percentage of the possible falls negative cases. 

 The field of deep technology has dramatically shifted during past 10 years  and the field of medicine is the one that may more than any other area of our life benefit from machine learning (ML)-powered tools. Today a lot of the consumers healthcare apps incorporated into the portable devices such as smartphones or smartwatches already use some kinds of the algorithms that track the physical condition of the user, store the data and create prediction about hearth attacks, strokes and diabetics complications. But in the clinical practice the level of using such kind of the technology solutions remains at the minimum level. This facts can be explained by the specific of the medical field which has a high level of safety requirements to the tools that might be used in the practical environment, as well as the strict requirements to the safety of the patients personal data. From the other side, the problem is that the number of the effective solutions available to the clinics right now are too small and usually this solutions are tailored specific for the need of the particular hospital and cannot be applied in the other clinics. This gap between the potential to use technological solution in the  diagnoses and the current politics of the hospitals affects negatively both the patients and practical specialists, as the work pressure for the doctors remains too high and the speed of the patients diagnose remains to low.

\subsection{Contribution}
The huge pressure and work-overload that faced medical specialist in the last year showed the urgent need in development the computer aided diagnose systems that will significantly reduce the human biased errors and help medical specialist create the prioritization lists for the emergency patients. Despite the big progress in the field of computer vision in the medical area, some branches of the medical diagnostics still lack the competent CAD solutions for the purpose of the diagnostics and further treatment. Our team applied the current state-of-art technology to the one of the most difficult and human-biased field of radiology - the problem of bone abnormalities detection using standard X-ray images. Using the proposed network architecture we have gathered good experiment results; this indicates that the further development of the CAD system for bone abnormalities detection based on the described in the article technology approach will lead do the shifting the quality of the real-time diagnostics.

\section{Previously works}
Because of the shared interest of the academia and practical specialists, the field of the medical ML solutions is rapidly growing for the last 10 years. The quality of the algorithm predictions has significantly raised and some of the areas of expertise are performed today better by algorithms rather by human specialists. One of the particular area, when today algorithms are outperforming practical specialists is image-based diagnoses, for example, ML algorithms detects spotting malignant tumours  better than experienced radiologists and can guide doctors how to construct cohorts for costly clinical trials \cite{Thomas}. Such fields of medicine as oncology, lung radiology, cardiology and brain surgery are already benefited from advanced Ml solutions, that used in the practical diagnostics\cite{Boon_2018} \cite{James_H}.  However, the better diagnoses is not the only issue that can be solved using algorithms, foe instance, optimizing the work load to prioritizing of the most difficult and urgent cases, pre-analyses of the cases in the high-loaded circumstances, where human biases such as a fatigue can lead to the negative effects and improving the quality of the medical images - are the areas that can get a significant support from the ML-tools incorporation\cite{James_H}.

It must be mentioned, that the area of the image-based diagnoses has largely benefited from rapid development of the computer vision technology, for instance, such initiatives as ImageNet and Kaggle helped to shift the image recognition to the new level. But a lot of image recognшtion medical tasks remained unsolved today. The field of trauma diagnostic is one of the example of the medical branch that relies heavily on the image-based diagnostics and it is one of the most promising directions of the ML-powered healthcare solutions development. But today the CAD technology in trauma radiology still haven't reached the acceptable level of accuracy, which would allow to use algorithms in the practical diagnosis. 

There is a substantial amount of work on the analysis in the field of computerized decision support systems for trauma patients. Thus, the main problem that remains is still a lack of the quality data. However some good results even with the small amounts of the training data have been achieved so far. For example, in his work Gene Kitamura reached 76 per cent accuracy of the ankle fracture predication using a small dataset (a total of 596 normal and abnormal ankle cases) \cite{Kitamura}. But using the bigger datasets can lead to even better results, that's why using the additional sources of the trained data as well as usage of the augmentation approach to the dataset enlargement become the crucial steps in developing the algorithm for CAD. For instance, Takaaki Urakawa reported the accuracy of 95.5 per cent in detection of the intertrochanteric hip fractures using the data from 1773 patients, who were enrolled in this study\cite{Urakawa}. Good results were also showed in detection of the distal radius fractures - using the dataset of 4,476 augmented anteroposterior hands X-ray images Erez Yahalomi and colleagues have achieved 96 accuracy in the radius bone fracture prediction\cite{Yahalomi}. Chi-Tung Cheng used 25,505 limb radiographs gathered between January 2012 and December 2017 and 3605 PXRs gathered between August 2008 and December 2016 for the algorithm pretraining and training to achieve 91 per cent accuracy in the limb fracture prediction\cite{Cheng1}. 

However, the size of the available training data is not the only problem that connected with the radiographs. The quality of the images as well as picture size requirements can differ significantly even within one country. The attempt to reduce the X-ray radiation doses affects the quality of the images and so called quantum noise appeared on the images, making them blurry and difficult to understand both for human-specialists and CAD \cite{TilAach}. Moreover, the absence of the universal requarements to the image quality also makes the diagnose process harder - images can be grayscale or color, high-resolution or low-resolution, hard copy or soft copy, uncompressed or compressed. All these difference in the radiographs caused up to 30 per cent error rate in the diagnosis every year, that lead to the complications in the patients treatment\cite{Krupinski}. So, another direction of the research and development with a great potential for the healthcare field is the area of image processing, which can significantly shift the quality of the X-ray images both for the needs of human doctors and intelligent systems. 

The other serious drawback of the current datasets is the absence of the universal approaches to the radiographs among different hospitals and private practical centers.  The difference in the image specifications that used in different clinics even withing the territory of one city makes the process of CAD implementation very difficult. To achieve better results additional steps such as analysis and processing of the bone fracture images should be used, however; there is different image issues that cant be solved even after the image processing steps. 

Besides the size of the training data, the technology and chosen architecture used in the study can affect the achieved level of accuracy. The other crucial factor of the study results is the type of CNN that were used in the CAD. For example, using RestNet50 for the relatively small dataset of the femure bones fractures showed a good result of 90 per cent accuracy for the fracture prediction \cite{JimenezSanchez}. Using the DensNet architecture for the hip fractures diagnosis showed also a promising results both on the small and large datasets- more than 90 per cent accuracy in the fracture detection using X-ray images\cite{Cheng}, \cite{Krogue}. InceptionV3 architecture was used to detetct fractures on the wirst and femur bones with relatively small data sets (1389 images) and showed a good result during the testing - 0.95 AUC \cite{Kim}.

The great diversity of the different metrics used to track the progress of the neural networks architecture described in the studies in the field of musculoskeletal radiographs abnormality detection should also be mentioned. Among the most widespread metrics of success that use different authors is the level of accuracy of prediction \cite{Chung}, \cite{Olczak}, \cite{Sher_Ee_Lim}. The other popular measurements of the fracture detection success are sensitivity (the percentage of true positive cases to the general amount of the positive predictions) and specificity (the percentage of true negative cases to the general amount of the negative predictions) \cite{Lindsey}, \cite{Cheng},\cite{Chung}. As for the competitions with MURA dataset the main index of the success of the algorithm is the Cohen’s kappa statistic, which expresses the agreement of the model with the gold standard. The use of Cohen’s kappa statistic is determined by the ability to provide more valuable information in  the  case  of  musculoskeletal  studies \cite{Saif}. Today the best showed Kappa score is 0.843, that was achieved in November 2018 year.

\section{Dataset}

As the source of the train and test data for our experiment MURA dataset of the musculosceletal abnormalities was used. MURA is one of the largest public radiographic image datasets published in Jan 2018 by Stanford ML Group. MURA dataset consists of musculoskeletal radiographs of 14,982 studies collected from 12,251 patients, with a total of 40,895 multi-view radiographic images. Each radiographs belongs to one of seven standard upper extremity radiographic study types: elbow, finger, forearm, hand, humerus, shoulder, and wrist. All images have standard format of 8-bit png. Each study was manually labeled by board-certified radiologists from the Stanford
Hospital at the time of clinical radiographic interpretation in the diagnostic radiology environment between 2001 and 2012. All the images were divided into the positive (with abnormality) and negative (without abnormality) cases and present the binary classification problem; no additional labeling from the practical specialists from our side on this stage was used due to the aim of the experiment.

To achieve better result through the diversity of the training and tested data we used data augmentation, for the purpose of this stage the next steps were made: random horizontal flip, random image rotation up to 30 degrees,image scaling randomly in the range of 95-130 per cent, brightness changing in the range 80-120 per cent. 


\section{Model}
 \subsection{Image pre-processing}
 
 MURA dataset of the bone abnormalities consist of the large variety of the different pictures with the different formats and sizes. Thus, one of the first problems that our team has faced working with this dataset was the necessity to make the data more homogeneous.  That's why for the purpose of the research work we used image pre-processing tools to achieve better quality of the images. The first step that we have made was to find the RoI in the image by binary threshold as well as to find contours of the image. This procedure allowed us to detect the region of the interest for the classification and to crop only RoI.
 
Data augmentation as an important part of the algorithm learning was also used during our research stage. To enlarge dataset we used different data augmentation approaches such as:
\begin{itemize}
  \item Random flipping the images horizontally;
  \item Random rotation of the images up to 30 degrees;
  \item Scaling randomly in the range 95-130 per cent;
  \item Randomly changing brightness in the range 80-120per cent;
\end{itemize}

Before using radiographs to the first step of the learning session they were normalized to have the same mean and standard deviation  of  images  in  the  ImageNet  training  set.

After the data augmentation normalization was used to DenseNet indexes of mean and std. This step was made to allow us change values of mean and std on the next stages of our project.

 \subsection{Network architecture and training}
 
The first step of our work was to choose between different architecture of Deep learning models to find the most appropriate one for the purpose of our project. The most widespread approaches for the image detection in the medical file is the usage of Depp Neural Networks (DNN)  and Convolutional Neural Networks (CNN). Thus, both of them showed pretty good results in the previous works, we decided to use CNN architecture based on two main chose criteria: good performance of the architecture with fast model training and the possibility to eliminate error propagation during the process of the learning sessions. As a result, 169-layer  convolutional  neural  network  was used to predict the probability of abnormality for each radiograph in MURA dataset.  The specifics of the chosen network is that it uses a Dense Convolutional Network architecture to connect each layer to every other layer in feed-forward fashion to make the optimize the deep networks results.    
For the purpose of the experiment the weights of the network were initialized with weights of the ImageNet pretrained model.
Taking into account the specifics of the MURA dataset and importance of the precise results for the medical diagnostics, during the learning sessions the batch size was chosen to be 8 images per session. Learning rate was 1e-4 -> and it was multiplied every 5-7 epochs by 0.1.
In average the number of epochs was  35-50 from ImageNet pretrained DenseNet-169. In the first pretraining session denselayer with freezed encoder was used (for first 5 epochs), then the whole network was unfreezed and it was trained 20-30 epochs more.

\begin{figure} [h]
    \centering
    \includegraphics [scale=0.3]{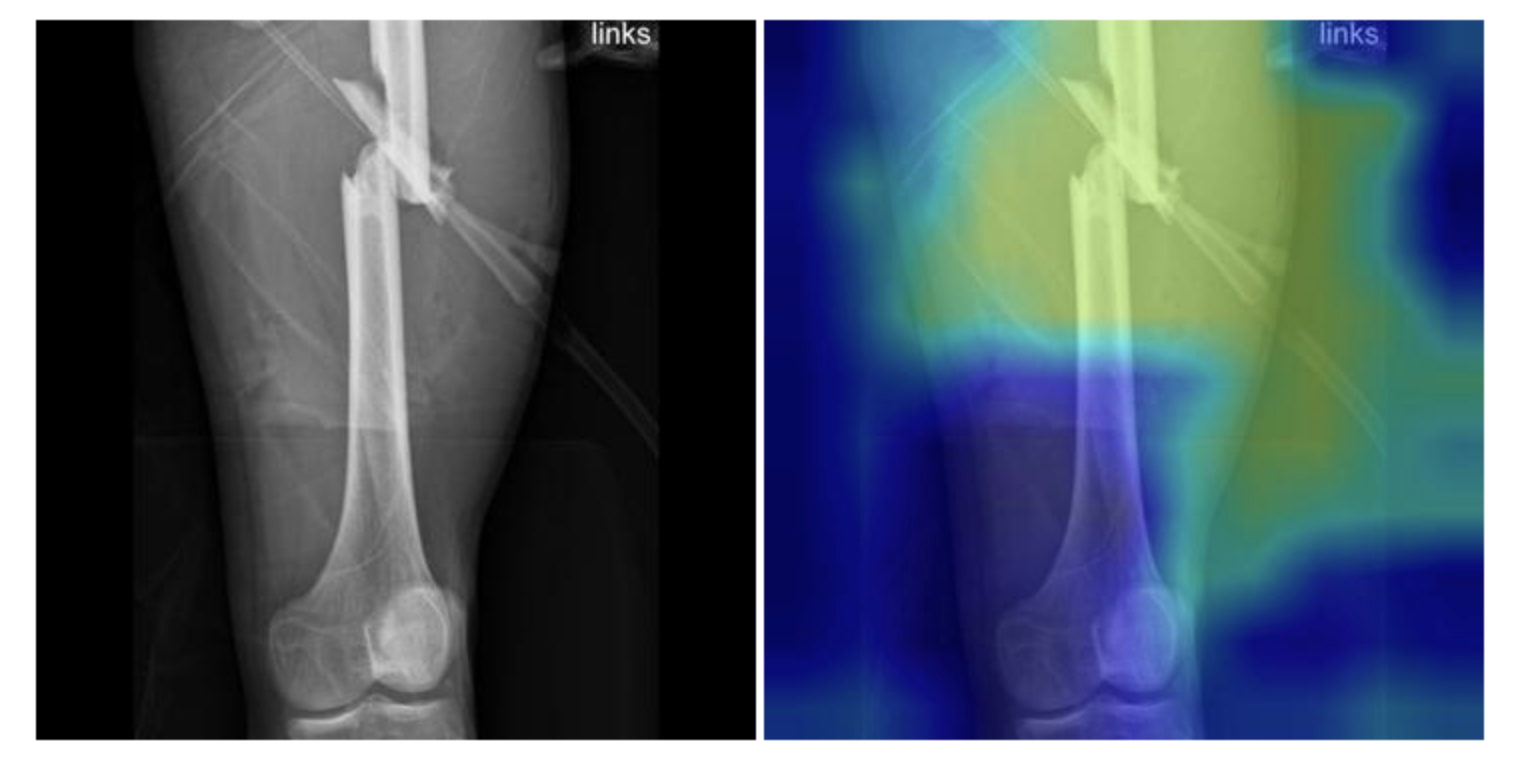}
    \caption{Example of the fracture detection on the leg bone}
    \label{fig:my_label2}
\end{figure}

\section{Results}

To understand the level of success of the proposed architecture normally the a new proposed system is compared to the already existing solution with the similar functionality. However, there are not so many working in the practical conditions bone abnormalities detection systems, that why we decided to compare achieved score with the results that were achieved in the similar studies.

As the measure of success of our solution we decided to chose three basic parameters - Kappa score, AUC ROC score and accuracy of the prediction. Kappa score as the main measurement index of the MURA dataset challenge was included to compare the achieved results with the results that were achieved directly on this data. The best results based on the Kappa score index that were achieved  by own solution belongs to the wrist (0.942), hand (0.862) and shoulder (0.735) radiographs. Thus, the result of our solution is slightly better than the maximum indexes showed on the official challenge web-site (0.931, 0.851 and 0.729 accordingly). AUC ROC has the index of 0.870 - this result showed the positive progress in the algorithm prediction, however, some issues connected with the image quality and diversity of the clinical conditions presented into the dataset remains that need to be solved to achieve a higher results. The accuracy of the prediction was estimated of  0.863, which is also a pretty good result for the CAD that were trained only using MURA dataset, but we see a lot of rooms for further improvement that we plan to work with on the next stages of our RnD work.

Graphical representation of the achieved results you can see on the Pic.1 and Pic. 2

 \begin{figure} [h]
    \centering
    \includegraphics [scale=0.4]{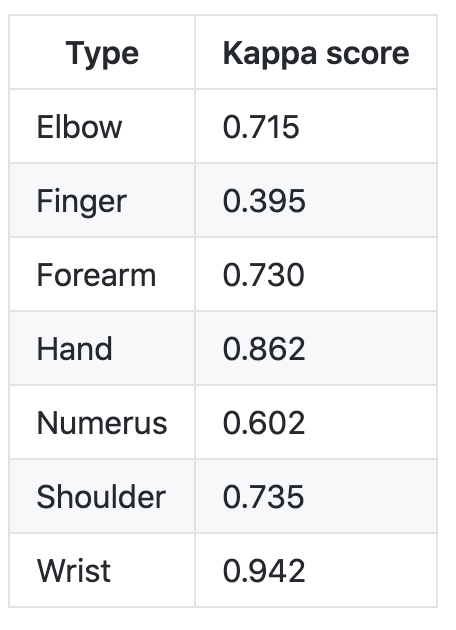}
    \caption{Prediction over study using 4 densenet-169 models ensemble (Cohen’s kappa statistic)}
    \label{fig:my_label}
\end{figure}

\section{Further Work}

Despite the good results achieved on this stage of our research project there are however some possibilities to increase the accuracy of the algorithm. The first of them is the further improvement of the data used for training the model. The quality and diversification of the data presented in MURA dataset to some degree affects negatively the results of the model training. The deep learning technique as the one that rely heavily on data, is very sensitive to the quality of the training data. All the radiographs presented in MURA dataset were made in clinics from 2001 to 2012 with the resolution of 600x600 pixels. The low resolution of the images almost eliminates the possibility to detect tiny fractures without precise labeling from the radiologists with a big practical experience. For instance, the current state of the radiograph resolutions used in the clinics is an average of 2000x2500, which is far more better that the ones used in MURA dataset \cite{Huda}.  To achieve the better results on the next stages of our work we plan to use radiographs with the current standard matrix-size collected from the last three years by the local clinics.  

The volume of the training data affects also the level of the accuracy that the model can achieved and solve one of the main factors of the project success. Thus, the MURA dataset present quite a big amount of the data in general, it gives not so many data to train the algorithm to train on the separate body parts. To gain the better result the train dataset should be no less than 10 000 samples, this will allow to get precise forecast about the fracture localisation. Pic 2 reflects the proportion of the cases in the MURA dataset according to the different parts of the body. As we can see on the diagram the number of the cases for each segment is far too small in comparison to the data needed to archive the credible results. 

\begin{figure} [h]
    \centering
    \includegraphics [scale=0.4]{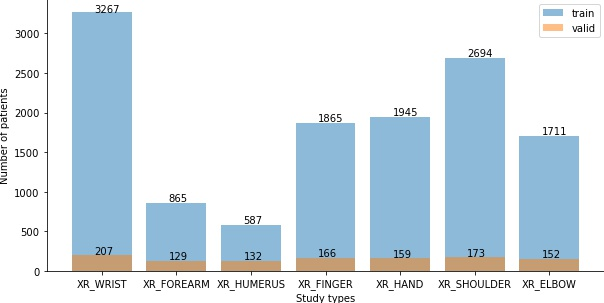}
    \caption{Number of the case studies according to each type of the fracture}
    \label{fig:my_label1}
\end{figure}

The other problem that we faced during this stage is the big variety of different clinical cases. Thus a great diversity of different types of the abnormalities in the combination with the list of different body parts used in the dataset creation can be a cause of the вecrease in success rates of the algorithm. On the next stages of our research work we plan to focus on the one type of the abnormality (fractures) and narrow down the dataset to only one type of the bones (arm). 

The absence of the labeling from the practical specialist is also an issue we plan to eliminate on the next stages. Extension of the initial information from a simple binary classification to the determination of a site with a specific bone abnormality will increase the accuracy of the abnormality prediction. To the purpose of this issue we made an agreement with the clinical experts with the average practical experience of 10 years that will work on the labeling of the radiographs for the next stage of our work.

\section{Conclusion}

To conclude, this article presents bone abnormalities detection model trained on the Stafford's MURA dataset. Muskoskelital abnormality detection is an important phase of the clinical diagnoses, that can be affected by the human biases such as tiredeness, big workload, lack of expertise, difficult-to-see cases of the bone fractures and muscoskeletal abnormalities, overworking, lack of concentration and multitasking of the radiologist.To eliminate this factors and to make the abnormality detection process faster and more reliable computer-aided diagnose systems should be incorporated into the working tools of the practical specialist. Ability to identify a zone of the abnormality  and further building of the prioritization list of the patients according to the degree of the danger of the possible diagnoses will significantly increase such crucial factor as the speed of the speed of medical care and will decrease the unnecessary work load for the doctors. However, such computer aided systems cannot replace radiologists in any way, but will significantly shifted the quality of their work. 

Our solution has showed a good results on the bone abnormalities detection, however it was focused only on the binary problem classification (existence or absence of the abnormality). To make it useful for the practical experience the model should be further trained to be able to distinguish different types of abnormalities with the stable high level of the accuracy.

It should also be mentioned, that besides the technological difficulty of the CAD system development, there is also an ethical problem, that directly affects the further development of this field. Despite the fact that ML-powered tools significantly reduce the work load for the practical specialists, as well as they increase the speed of the diagnoses, the ethic aspect of using algorithms to predict some deviations in the health condition of the patient remains a big issue that directly affects the incorporation these solutions into the daily practice of the radiologists. There are two main approaches in developing ML-based medical applications - supervised and unsupervised learning. An while, the first one is more simple an understandable for human-operators, the second one is more promising, as it is the one that can significantly reduce the human biases into the algorithms work and, thus, shift the accuracy of the prediction. But the mechanism of unsupervised algorithm can be very hard to understand and in the healthcare branch, as the one of the most regulated ones, its almost impossible to get the permission for its usage in the clinical conditions. This means that further steps to the cooperation between medicine and technology should be made to make the research projects a part of the clinical practice.

\bibliographystyle{unsrt}       
\bibliography{literature} 

\end{document}